\documentclass[preprint,aps,prb,showpacs,amsmath,amssymb,longbibliography,nobibnotes]{revtex4-1}

\pdfoutput=1

\usepackage{graphicx}			
\usepackage{bm}           
\usepackage{natbib}       

\usepackage{ifpdf}

\ifpdf
  \usepackage[pdftex]{hyperref}
  \hypersetup{
    colorlinks=false,
    pdfpagemode=UseThumbs,
    plainpages=false
   }
\else
  \usepackage[dvips]{hyperref}
  \hypersetup{ 
    colorlinks=false,
    pdfpagemode=UseThumbs,
    plainpages=false
   }
\fi

\begin{document}

\title[Quantum Hall Resistance Overshoot in 2DEG - Theory and Experiment]{Quantum Hall Resistance Overshoot in 2-Dimensional Electron Gases - Theory and Experiment}

\author{J.~\surname{Sailer}}
\affiliation{Walter Schottky Institut, Technische Universit\"at M\"unchen, Am Coulombwall 3, 85748 Garching, Germany}

\author{A.~\surname{Wild}}
\affiliation{Walter Schottky Institut, Technische Universit\"at M\"unchen, Am Coulombwall 3, 85748 Garching, Germany}

\author{V.~\surname{Lang}}
\affiliation{Walter Schottky Institut, Technische Universit\"at M\"unchen, Am Coulombwall 3, 85748 Garching, Germany}

\author{A.~\surname{Siddiki}}
\email{asiddiki@fas.harvard.edu}
\affiliation{Istanbul University, Physics Department, Faculty of Sciences, 34134 Vezneciler, Istanbul, Turkey}
\affiliation{Harvard University, Physics Department, 02138 Cambridge MA, USA}

\author{D.~\surname{Bougeard}}
\email{bougeard@wsi.tum.de}
\affiliation{Walter Schottky Institut, Technische Universit\"at M\"unchen, Am Coulombwall 3, 85748 Garching, Germany}
\affiliation{Institut f\"ur Experimentelle und Angewandte Physik, Universit\"at Regensburg, 93040 Regensburg, Germany}

\date{\today}

\begin{abstract}
We present a systematical experimental investigation of an unusual transport phenomenon observed in two dimensional electron gases in Si/SiGe heterostructures under integer quantum Hall effect (IQHE) conditions. This phenomenon emerges under specific experimental conditions and in different material systems. It is commonly referred to as Hall resistance overshoot, however, lacks a consistent explanation so far. Based on our experimental findings we are able to develop a model that accounts for all of our observations in the framework of a screening theory for the IQHE. Within this model the origin of the overshoot is attributed to a transport regime where current is confined to co-existing evanescent incompressible strips of different filling factors.
\end{abstract}


\maketitle

\section{Introduction}
\label{sec:L01_Intro}

Two-dimensional electron gases (2DEG) at low temperatures and high magnetic fields provide a foundation to investigate the properties of interacting many-particle systems. The observation of the integer quantum Hall effect (IQHE) \cite{vKlitzing80:494} is one manifestation of the properties of such a 2D system and most of its essentials are successfully described within a single particle picture by means of the Landau-B\"uttiker formalism \cite{Buettiker86:1761}. Nevertheless, many anomalies of the IQHE which have been observed experimentally \cite{Ahlswede01:562,Yacoby04:328,Hor2008} cannot be explained in this framework.

One of these anomalies is the so called quantum Hall resistance overshoot in which the transverse resistivity $\rho_\mathrm{xy}$ exhibits an unusual behavior. Instead of monotonically increasing as a function of magnetic field $B$ between successive Hall plateaus, the Hall resistance overshoots a plateau at its low magnetic field end before dropping back to the respective Hall resistance value.

This feature has already been observed in various material systems like 2DEGs in GaAs/AlGaAs \cite{KOM1991,RIC1992,KOM1993}, Si/SiGe \cite{Wei1996,Gri2000} and GaInAs/InP \cite{Ram1998} heterostructures and Si-MOSFET samples \cite{Shl2005,Shl2006}. From these studies, two main approaches \cite{KOM1991,RIC1992} have been developed to explain the overshoot phenomenon. Yet, no model has emerged so far that describes the various characteristics of the overshoot consistently and independently of the specific properties of a respective material system.

In recent years, new light was shed on the origin of the IQHE by treating 2DEGs within a self-consistent screening theory \cite{siddiki2004,Gerhardts2008} that also takes direct Coulomb interactions into account. This theory is capable of explaining many experimentally observed details of the IQHE \cite{Ahlswede01:562,Yacoby04:328,Hor2008,Friedland:09,Mares:09}. The interpretation of the IQHE within the screening theory is in contrast to the classical Landauer-B\"uttiker based edge state. In the Landauer-B\"uttiker picture, current flows along independent and spatially separated 1D edge channels, whereas in the screening theory, the imposed non-equilibrium current is confined to one incompressible (edge-)state with a respective filling factor during the existence of a quantum Hall plateau.

We present a thorough investigation of the overshoot phenomenon based on 2DEGs in molecular beam epitaxy (MBE) grown Si/SiGe heterostructures for which we find the overshoot to be well developed and tunable by different experimental parameters, in particular the temperature, the 2D sheet carrier density, the sample current as well as the sample geometry. We develop a model for the overshoot phenomenon within the framework of the screening theory that turns out to be a natural outcome of current confinement to co-existing incompressible regions of different filling factor. Our model is universal to explain the Hall resistance overshoot over a wide range of different experimental parameters and confirms our experimental findings while it does not rely on peculiarities of the material system. Our work also illustrates the effect of current confinement to more than one edge state within the screening theory for the first time. 

The article is organized as follows: In section~\ref{sec:L02_ScreeningBasics}, we lay out the basics of the screening theory of the IQHE. After introducing the experimental setup and sample structure in section~\ref{sec:L03_SetupMethods} we report on the experimental characterization of this quantum Hall effect anomaly in section~\ref{sec:L03_Experiments}. Having identified the origin of this phenomenon, section~\ref{sec:L04_Discussion} explains the existence of the overshoot in the screening framework of the IQHE. In addition, the model is verified by comparing its implications with our experimental data and that of other groups.

\section{Screening Theory of the IQHE - Basics}
\label{sec:L02_ScreeningBasics}

At low temperatures and within certain magnetic field intervals, the screening theory gives rise to a fragmentation of the electron gas into incompressible and compressible spatial regions \cite{Chklovskii92:4026}. The 2DEG is called compressible, if the Fermi energy is pinned to one of the discrete levels. In this case, the electrons can easily redistribute due to a high density of states. In contrast, within the incompressible regions, all available states below the Fermi energy are occupied. Therefore electrons cannot be redistributed in the zero temperature limit. Hence, screening is nearly perfect (poor) in the (in)compressible regions. Note that, since all the states are occupied within an incompressible region, the filling factor is an integer. In this IQHE theory, the existence of an incompressible strip (IS) leads to the confinement of the entire current distribution to such an IS where no backscattering can occur. This current confinement leads to vanishing longitudinal resistance and quantized Hall plateaus \cite{siddiki2004}.

\begin{figure}
\includegraphics[width=1\columnwidth]{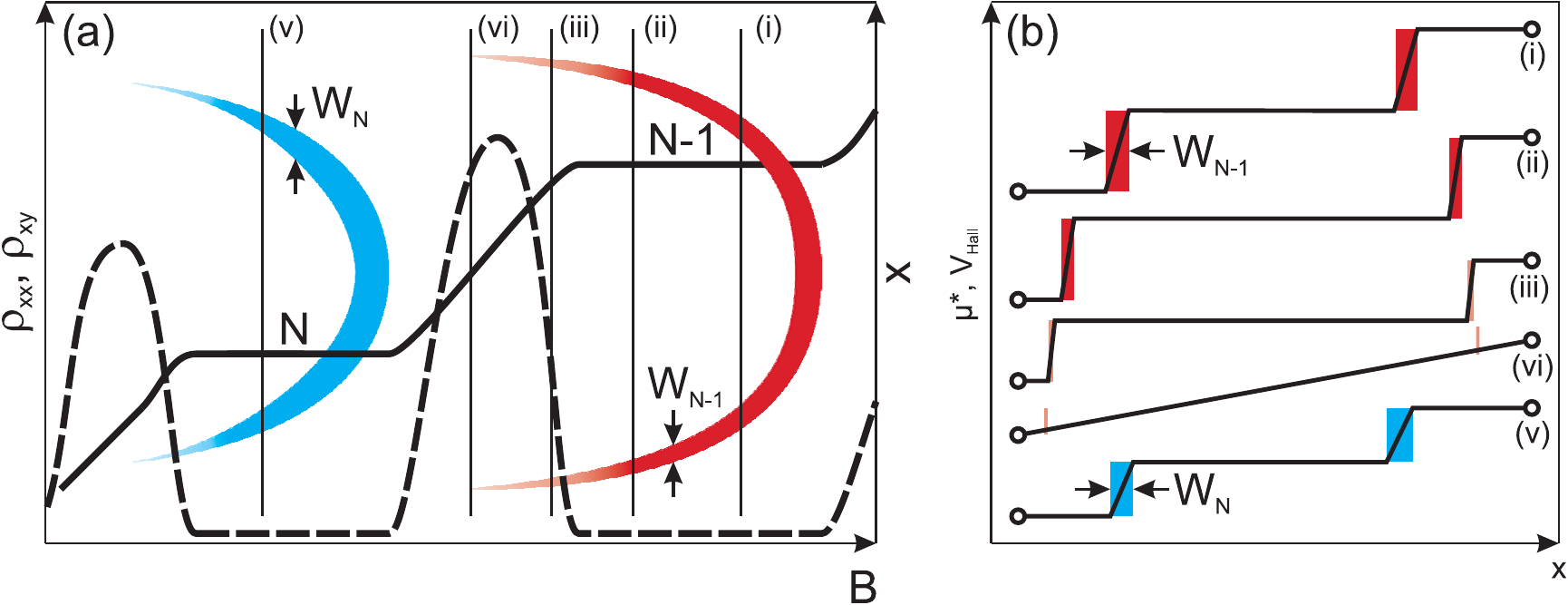}

\caption{(Color online) (a) Schematic longitudinal ($\rho_\mathrm{xx}$) and transverse ($\rho_\mathrm{xy}$) resistivity as a function of magnetic field. Crescent-like structures mark the course of incompressible regions of the 2DEG with magnetic field for filling factor $N-1$ (red) and $N$ (blue). Formation of IS regions correlate with the observation of a quantized Hall resistance. In the quantized regime, the imposed sample current is confined to the IS where no backscattering occurs. For very narrow \emph{evanescent} IS, $W<\lambda_\mathrm{F}$, current leaks out of an IS and quantization is lost. (b) Course of $\mu^*$ across the sample for different magnetic fields. For (i) and (ii), current is confined to the well developed IS resulting in a quantized drop of the Hall voltage $V_{\mathrm{Hall}}$. As an IS becomes evanescent (iii), current also flows in the compressible region and no transverse electric field $E_x\propto\partial_x\mu^*(x,y)$ is generated. Thus, the Hall plateau breaks down. For further decreasing magnetic fields, current spreads over the entire sample width (iv) and $\mu^*$ drops linearly across the sample, like in the classical Hall effect.}

\label{fig:01_ScreeningBasics}
\end{figure}

In order to introduce all relevant figures of merit that are required for the experimental section, figure~\ref{fig:01_ScreeningBasics} is used to review the existence of the IQHE within the screening theory. Figure~\ref{fig:01_ScreeningBasics}(a) shows a typical set of Hall traces for $\rho_\mathrm{xx}$ and $\rho_\mathrm{xy}$ in a magnetic field interval where two arbitrary quantum Hall plateaus are present. The overlayed red and blue crescent-like like structures highlight the transverse position $x$ across the Hall bar of IS regions for different magnetic fields. For a given magnetic field, the colored region is incompressible, whereas the surrounding (white background) is compressible. The existence of IS correlates with the formation of the quantized regime in the resistivity tensor. 

In figure~\ref{fig:01_ScreeningBasics}(b), the electrochemical potential $\mu^*$ is sketched versus the transverse position $x$ of the Hall bar for different magnetic fields. The traces (i) to (v) correspond to the vertical cross-sections as indicated in figure~\ref{fig:01_ScreeningBasics}(a). Such an evolution of $\mu^*$ for different magnetic fields has also been observed experimentally \cite{Ahlswede02:165} and described theoretically \cite{Siddiki:ijmp}. The experimentally observable difference of $\mu^*$ on both sides will be the Hall voltage $V_\mathrm{Hall}$. 

Starting at the high magnetic field end in figure~\ref{fig:01_ScreeningBasics}(a), the entire sample is compressible and consequently behaves like a metal. Hence, a linear drop of $\mu^*$ across the sample and a linear variation of the Hall resistance versus magnetic field is observed as in the classical Hall effect. Lowering the field strength results in formation of an IS in the center of the Hall bar with a local filling factor $\nu(x)=N-1$ and $\rho_\mathrm{xy}$ becomes quantized. We will denote the center of the 2DEG \emph{'bulk'} from now on. Due to the absence of backscattering in the IS, current will flow in the IS only. By lowering the magnetic field, at some point the bulk of the 2DEG becomes compressible again. The incompressible region in figure~\ref{fig:01_ScreeningBasics}(a) splits and two branches of the IS develop that approach the sample edges when further decreasing the magnetic field. This is illustrated in the cross-sections (i) and (ii): as both IS branches move to the sample edges, they become narrower, however, current is still confined to the IS as long as the sample is in the quantized regime. This condition is fulfilled if the widths $W_{N},W_{N-1}$ of these IS are larger than both, the mean spacing between charge carriers, as well as the spatial extent of the wave function. While the first scale is described by the Fermi wavelength $\lambda_\mathrm{F}\propto\sqrt{1/n_{\mathrm{2DEG}}}$, the latter is on the order of the magnetic length $l_B=\sqrt{\hbar/eB}$.

Below a certain width $W_{N-1}$ of the IS, quantization will be lost and the IS breaks down for two possible reasons: In the first case, if $l_B<W_{N-1}<\lambda_\mathrm{F}$, current starts to leak out of such an IS into the surrounding compressible region. In such a case, the mean distance between charge carriers is larger than the width of the IS. Thus, it is statistically unlikely to find any charge carriers within the IS region and the strip cannot be defined as an incompressible state anymore. Current outside an IS is not quantized due to finite scattering \cite{Siddiki04:condmat}, and as a result, the Hall resistance is reduced. This situation is depicted in cross-section (iii). We will call the IS in this regime \emph{evanescent} from now on. In the second case, for $W_{N-1}<l_B<\lambda_\mathrm{F}$ at even lower magnetic fields, the width of the evanescent IS will become smaller than the extent of the wave function. Consequently, electrons will be able to tunnel through the IS and the current will spread over the entire Hall bar, such that classical Drude-like transport is recovered. In cross section (iv), $\mu^*$ will not only drop in the former IS regions but linearly across the entire sample and the Hall resistance continues to decrease. The above described scheme perfectly agrees with local probe experiments \cite{Ahlswede01:562,Ahlswede02:165}. If the magnetic field is chosen such that $\lambda_\mathrm{F}<l_B$ at the low magnetic field end of a Hall plateau, the first case does not occur and the IS breaks down immediately. By decreasing the magnetic field further, the bulk becomes incompressible for filling factor $N$ and current is confined to the IS again as illustrated in cross-section (v).

This pictorial example is valid for low injected currents that do not strongly distort the electrostatics of the problem, as in the out-of-linear-response regime that has been discussed already within the screening theory \cite{Guven03:115327,SiddikiEPL:09}.

From this theory of the IQHE it becomes clear that the existence of a Hall resistance plateau is directly linked to the existence of one IS that carries the imposed current in a quantized number of levels. At the low magnetic field end of a Hall plateau, the existence of an IS itself is determined by its width, position and the energetic stability of the charge and current distribution. These characteristics are in turn related to the respective energy gap \cite{afifPHYSEspin}, the electron density gradient at the sample boundaries \cite{Chklovskii92:4026,Sefa08:prb}, the strength of the magnetic field \cite{siddiki2004}, the temperature \cite{Lier94:7757,Oh97:13519} and the amplitude of the imposed current \cite{Akera06:}. Thus, these physical dependencies define experimental parameters with which the existence of IS and possible anomalies of the IQHE can be studied.

\section{Samples and Methods}
\label{sec:L03_SetupMethods}

\begin{figure}
\includegraphics[width=1\columnwidth]{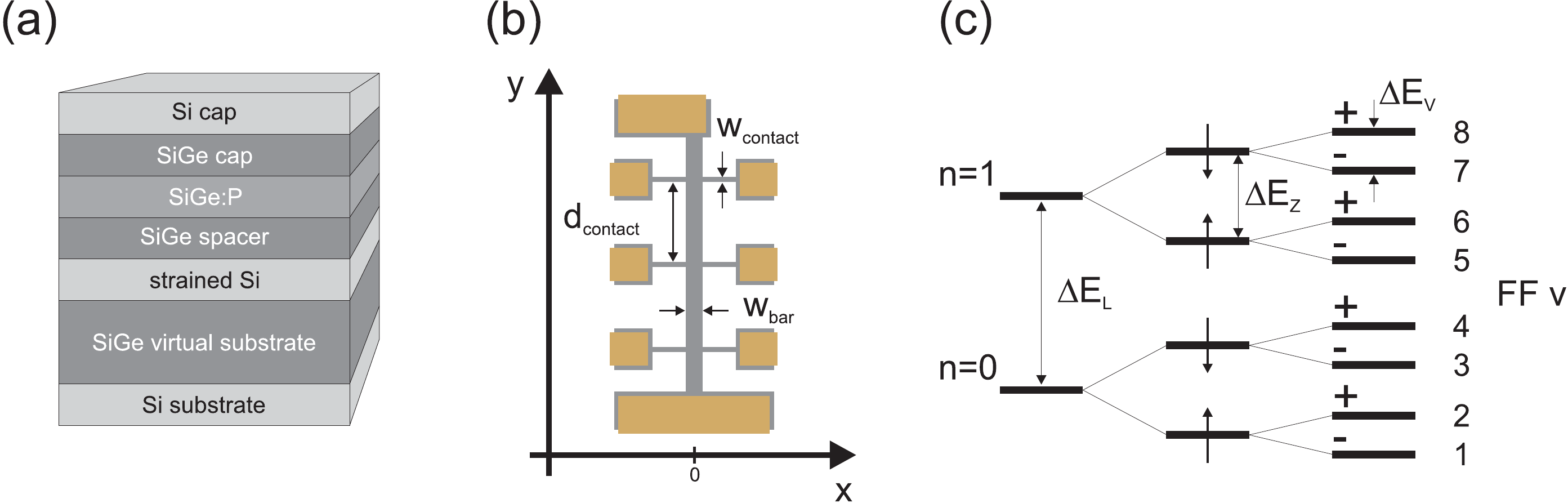}

\caption{(Color online) (a) Sketch of the layer structure of Si/SiGe heterostructures. (b) Hall bar layout with markers for the relevant dimensions $w_{\mathrm{bar}}$, $d_{\mathrm{contact}}$ and $w_{\mathrm{contact}}$. (c) Energy level scheme for 2DES in strained Si subject to a perpendicular magnetic field.}

\label{samplelayout}
\end{figure}

The sample layout is shown in figure~\ref{samplelayout}(a). Samples are fabricated in a Riber Siva 45 solid source molecular beam epitaxy machine. Growth is initiated on a $\mathrm{1500~{\Omega}\,cm}$ (100) oriented Si substrate. The virtual substrate necessary to induce the biaxial tensile strain in the Si, is grown employing either a graded buffer virtual substrate concept with a grading rate of $\mathrm{8\%/{\mu}m}$ or a low-temperature silicon virtual substrate \cite{Che1996,Lin1997,Li1997a}. The 2DEG forms in the $\mathrm{15~nm}$ thick strained Si layer which is followed by a $\mathrm{15~nm}$ SiGe spacer layer and a $\mathrm{15~nm}$ SiGe:P dopant supply layer. The uppermost $\mathrm{45~nm}$ SiGe layer and the $\mathrm{10~nm}$ cap layer combined with the doping concentration account for the necessary band bending due to Fermi level pinning at the surface and protect the structure against oxidation. All heterostructures studied in this contribution have typical mobilities of $\mathrm{1-2\times10^4~cm^2\,(V\,s)^{-1}}$ at electron densities of $\mathrm{3-4.5\times10^{11}~cm^{-2}}$. 

The energy level structure of such 2DEGs is schematically shown in figure~\ref{samplelayout}(c) for an applied perpendicular magnetic field. Landau levels are separated by the largest energy gap ${\Delta}E_\mathrm{L}=\hbar\omega_\mathrm{c}$. Each Landau band splits into two Zeeman levels separated by ${\Delta}E_\mathrm{Z}=g\mu_\mathrm{B}B$ with an electron \emph{g}-factor of 2 in Si \cite{Gra1999}. In the strained Si two-valley system, both valley bands are split by ${\Delta}E_\mathrm{V}={\Delta_\mathrm{V}}B$ which is typically on the order of $\mathrm{20~{\mu}eV/T}$ \cite{Goswami2006}. The various energy gaps can only be resolved individually in magneto-transport measurements if the respective energy splitting exceeds the level broadening $\Gamma$, \emph{i.e.} $\Gamma\ll{\Delta}E_\mathrm{L,Z,V}$.

Electrical characterization of the samples is carried out in Hall bar geometry at temperatures down to $\mathrm{320~mK}$ and magnetic fields up to $\mathrm{10~T}$ in a $\mathrm{^3{He}}$ cryostat. Hall bars are defined using photolithography and wet chemical etching. Figure~\ref{samplelayout}(b) sketches a typical Hall bar layout and names relevant length scales. The width $w_\mathrm{bar}$ of the Hall bars is varied between $\mathrm{20~{\mu}m}$ and $\mathrm{200~{\mu}m}$. The distance $d_\mathrm{contact}$ between adjacent contacts is chosen always to exceed $w_\mathrm{bar}$ by a factor of $\mathrm{10}$. Measurements are performed by using a standard low-frequency lock-in technique at $\mathrm{17~Hz}$, voltage preamplifiers with $\mathrm{1~T\Omega}$ input impedance \cite{Fis2005} and an excitation current of typically $\mathrm{10~nA}$ to avoid resistive heating of the 2DEG. In order to increase the 2D sheet carrier density, samples can be illuminated with a red light emitting diode at low temperatures.

\section{Experiments}
\label{sec:L03_Experiments}

In this section, a systematical investigation of the overshoot strength for various experimental parameters is discussed that were identified to influence the energetic stability of an incompressible edge state during the existence of a quantum Hall plateau.

\begin{figure}
\begin{center}
\includegraphics[width=0.6\columnwidth]{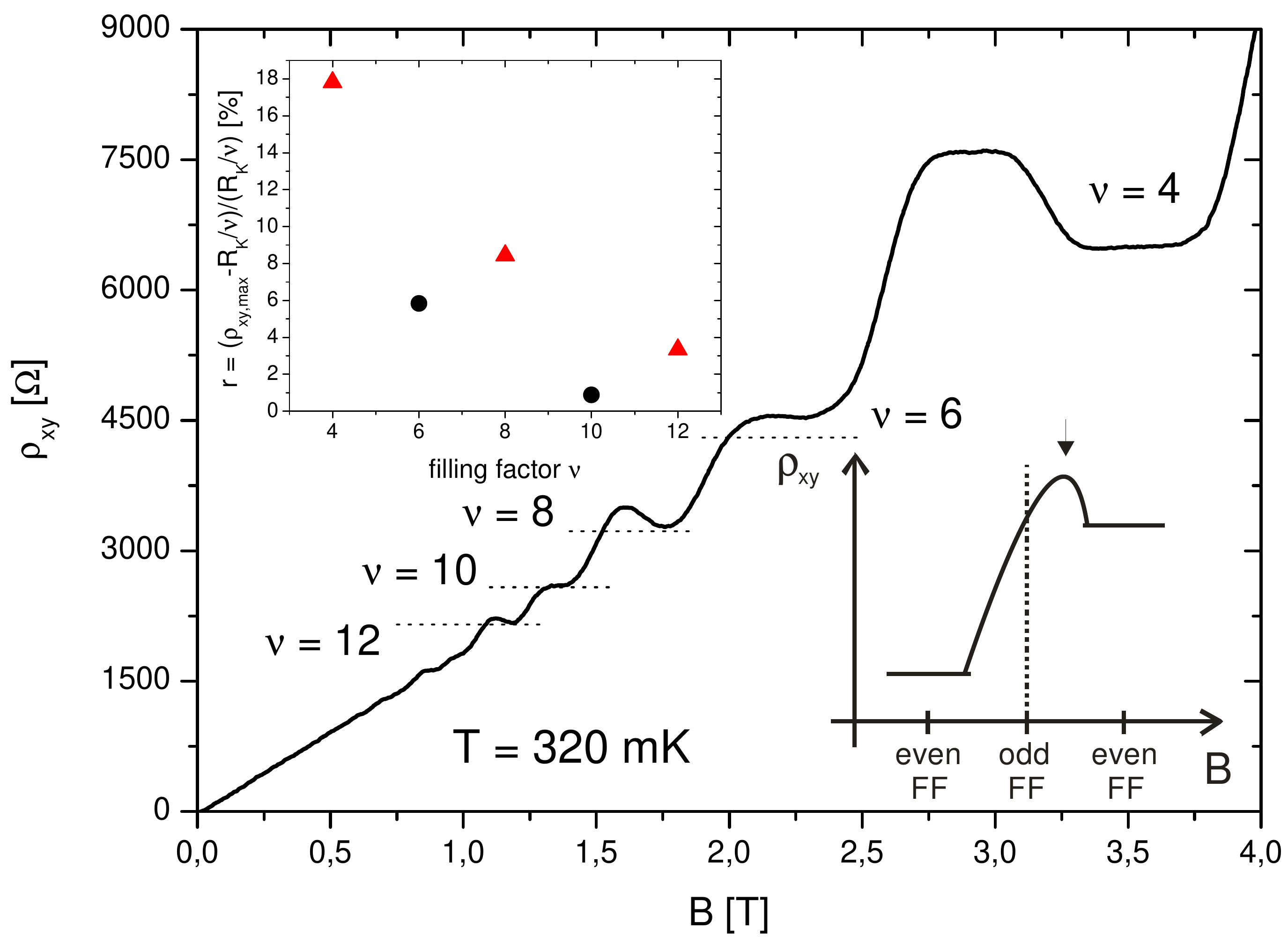}
\end{center}

\caption{(Color online) A typical Hall measurement of a sample showing an overshoot at filling factors $\nu$ 4, 6, 8, 10 and 12. Upper left inset: Relative strength of the overshoot at different filling factors. Lower right inset: Sketch of $\rho_\mathrm{xy}$. The maximum between two even filling factors is reached after the expected position of the odd filling factor in between.}

\label{Intro}
\end{figure}

Figure~\ref{Intro} presents a typical measurement recorded at $T=\mathrm{320~mK}$ using a excitation current of $\mathrm{10~nA}$ in a $w_\mathrm{bar}=\mathrm{20~{\mu}m}$ wide Hall bar. As expected, quantum Hall plateaus corresponding to the experimentally resolvable filling factors develop in the Hall resistance $\rho_\mathrm{xy}$. In contrast to well known quantum Hall traces, no monotonic behavior of $\rho_\mathrm{xy}$ is observed. Instead, at the low magnetic field end, $\rho_\mathrm{xy}$ exceeds, or overshoots the respective quantized plateau value of $\rho_\mathrm{xy}$. As a guide to the eye, the nominal values that equal $R_\mathrm{K}/{\nu}$ are marked with dotted lines, with the von Klitzing constant $R_\mathrm{K}=h/e^2\approx \mathrm{25812.8~\Omega}$.
The lower right inset of figure~\ref{Intro} sketches the general shape of $\rho_\mathrm{xy}$ valid for the observed overshoots. Although we do not resolve the odd filling factors, overshoots can be attributed to the even filling factors as the maximum of $\rho_\mathrm{xy}$ is always reached well beyond the expected position of the odd filling factor.

We ensured that the overshoot phenomenon is no artifact due to admixtures of $\rho_\mathrm{xx}$ in $\rho_\mathrm{xy}$. The effect is rather reproducible for different voltage probe contacts of a Hall bar under DC or low-frequency AC excitation, independent of sample processing, and shape and width $w_\mathrm{contact}$ of the probe contacts.

In our heterostructures, only even filling factors associated with Landau or spin gaps are resolved in Hall measurements. All of them display an overshoot. Due to the small valley splitting, odd filling factors cannot be resolved under the given experimental conditions. The upper left inset of figure~\ref{Intro} shows the relative strength of the overshoot $r=(\rho_\mathrm{xy,max}-\frac{R_\mathrm{K}}{\nu})/\frac{R_\mathrm{K}}{\nu}$ at the respective filling factor and reveals that overshoots are most pronounced for Hall plateaus of Landau gap associated filling factors at ${\nu}=\mathrm{4}, \mathrm{8}, \mathrm{12}, \dots$ compared to overshoots at Hall plateaus of smaller spin gap associated filling factors ${\nu}=\mathrm{6}, \mathrm{10}, \dots$. The energy gap size in turn, was identified to determine the stability of an IS \cite{afifPHYSEspin}.

\begin{figure}
\begin{center}
\includegraphics[width=0.6\columnwidth]{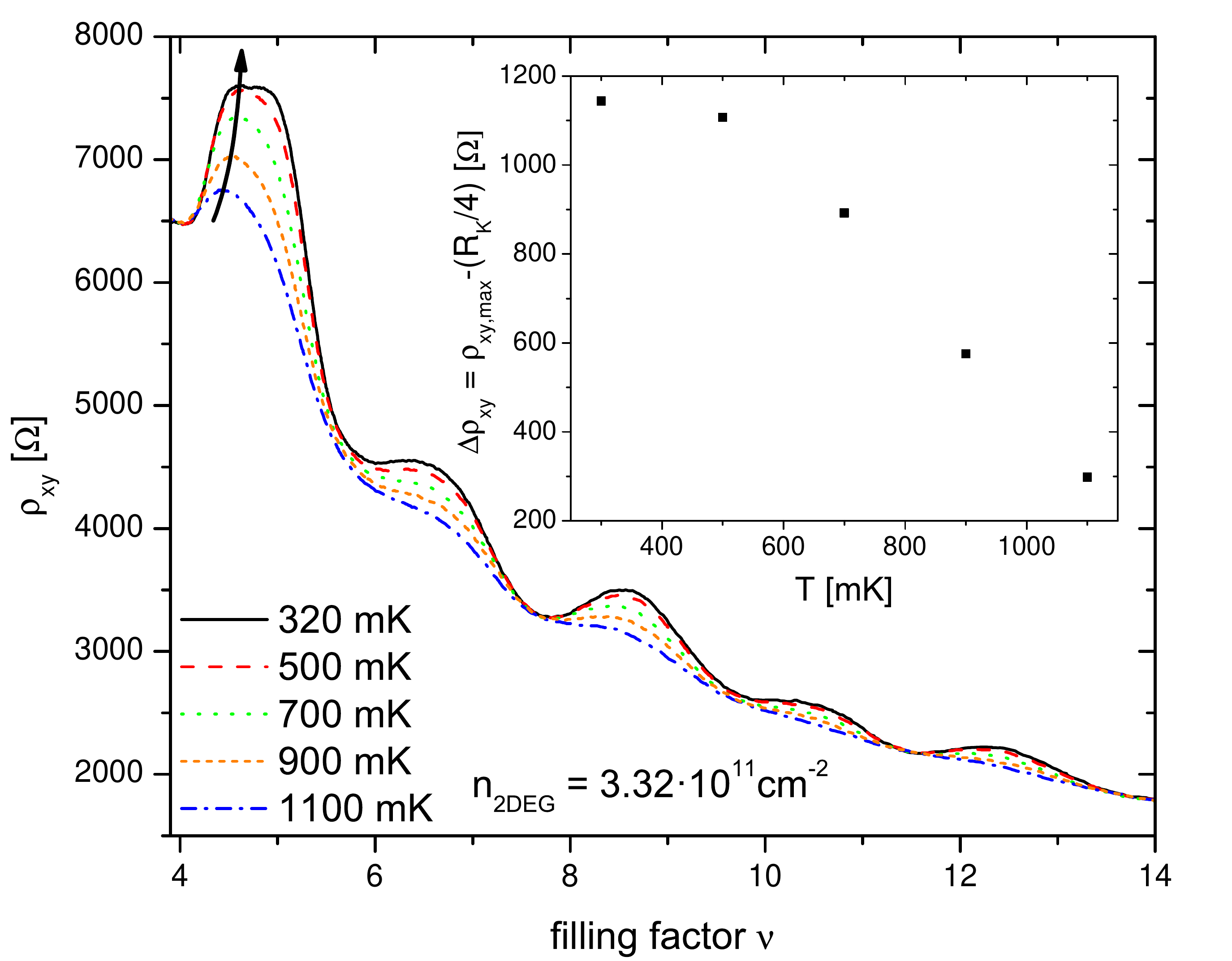}
\end{center}

\caption{(Color online) Temperature dependence of the overshoot. The overshoot increases with decreasing temperature. The inset shows the difference between the overshoot maximum at ${\nu}=4$ and $R_\mathrm{K}/4$ plotted versus temperature.}

\label{T-series}
\end{figure}

As the stability of an incompressible strip is strongly dependent on the electron temperature, we can also expect the temperature to influence the overshoot effect. Heating the electron gas leads to increased electron scattering and thus smears out the incompressible strip \cite{Oh97:13519}.
In figure~\ref{T-series}, $\rho_\mathrm{xy}$ is plotted as a function of the filling factor and illustrates the evolution of the overshoot phenomenon for a $w_{\mathrm{bar}} = \mathrm{20~{\mu}m}$ wide Hall bar when decreasing the temperature from $\mathrm{1.1~K}$ down to $\mathrm{320~mK}$. With decreasing temperature, the magnitude of the overshoot becomes increasingly prominent at all experimentally observed plateaus. Additionally, the position of the maximum of the overshoot shifts away from the associated Hall plateau towards higher filling factors with decreasing temperature. The position of the maximum of the overshoot at filling factor 4 is marked, as a guide to the eye, with a black arrow. Furthermore, the inset of figure~\ref{T-series} plots the difference $\Delta\rho_\mathrm{xy}=\rho_\mathrm{xy,max}-\frac{R_\mathrm{K}}{4}$ of the maximum of the overshoot at filling factor 4 and $R_\mathrm{K}/4$ as a function of the measurement temperature. For temperatures between $\mathrm{1.1~K}$ and $T\approx\mathrm{600~mK}$, we observe an increase of the magnitude of the overshoot. Below $T\approx\mathrm{500~mK}$, the overshoot saturates. This observed temperature dependence is in agreement with observations of Griffin \emph{et al.} \cite{Gri2000} in Si/SiGe 2DEGs, but deviates from observations of Komiyama \emph{et al.} \cite{KOM1991} and Richter \emph{et al.} \cite{RIC1992} in GaAs. 

\begin{figure}
\begin{center}
\includegraphics[width=0.6\columnwidth]{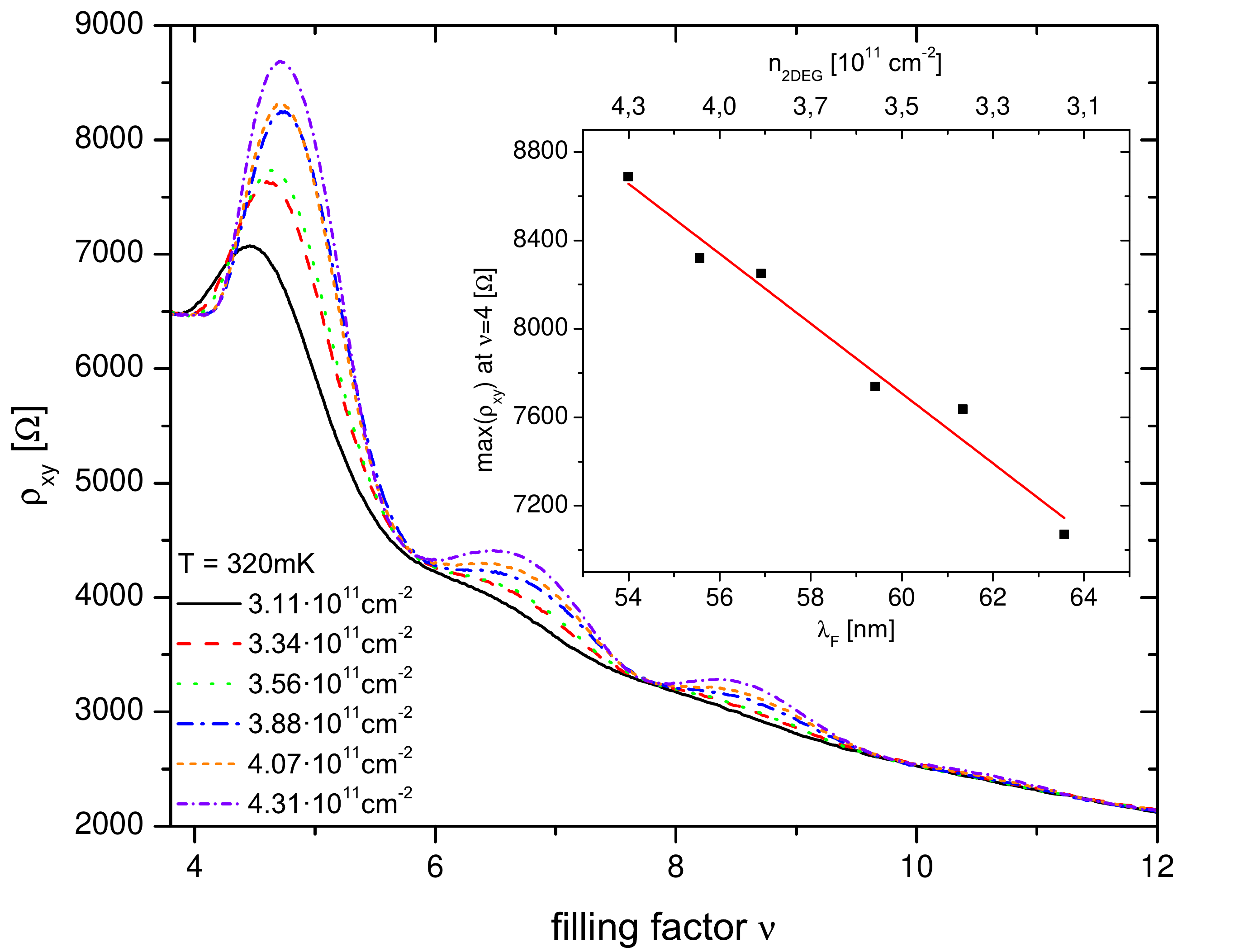}
\end{center}

\caption{(Color online) Dependence of the overshoot on the 2D sheet carrier density. 2D sheet carrier density was varied by illumination. Inset: $\rho_{xy,max}$ of the overshoot at ${\nu}~=~4$ increases with increasing 2D sheet carrier density, \textit{i.e.} decreasing Fermi wavelength $\lambda_\mathrm{F}$.}

\label{N-series}
\end{figure}

We now study the dependence of the overshoot on the 2D sheet carrier density. A varying density corresponds to a varying Fermi-wavelength $\lambda_\mathrm{F}$. In Si, $\lambda_\mathrm{F}$ equals to $\mathrm{2}\pi/{\sqrt{\mathrm{4}{\pi}n_{\mathrm{2DEG}}/{g_\mathrm{s}}g_\mathrm{v}}}$ with both spin- and valley-degeneracy factors $g_\mathrm{s}=\mathrm{2}$ and $g_\mathrm{v}=\mathrm{2}$. By changing the density of the 2DEG via illumination, we are thus able to tune the length scale below which an incompressible strip becomes evanescent as described in section~\ref{sec:L02_ScreeningBasics}. 
A low Fermi wavelength enables more narrow, however, still stable incompressible strips. Decreasing $\lambda_\mathrm{F}$ thus extends the magnetic field range in which an incompressible strip is able to carry current. An increase of the 2D sheet carrier density improves the screening behavior in the surrounding compressible region further and thus tends to suppress electron scattering.
In figure~\ref{N-series}, $\rho_\mathrm{xy}$ is plotted for six different electron densities. The overshoot is found to increase with increasing 2D sheet carrier density. Remarkably, for the highest density, the overshoot at filling factor 4 even exceeds $R_\mathrm{K}/\mathrm{3}\approx \mathrm{8604.3~\Omega}$ which corresponds to a valley-split energy level of the second spin band. Moreover, with increasing density, and similar to the temperature behavior shown in figure~\ref{T-series}, the overshoot shifts away from the quantum Hall plateau in direction of higher filling factors. Exemplarily for the overshoot at filling factor 4, the inset of figure~\ref{N-series} plots the maximum value of this overshoot versus $\lambda_\mathrm{F}$ and the corresponding sheet carrier density. We observe an increasing overshoot with decreasing $\lambda_\mathrm{F}$. 

Two main models have emerged to explain the overshoot phenomenon. In what we call the \emph{bulk model}, the overshoot is caused by backscattering in the bulk of the 2DEG due to energy level overlaps or crossings \cite{RIC1992,Wei1996}. In the second approach, which we will refer to as \emph{edge model}, the origin of the overshoot is attributed to a mixing of the two spin states of a Landau level at the sample boundaries and selective probing of only the outer edge channels by the voltage probe contacts \cite{KOM1991}.
The energy gap, temperature and density dependence, however, provide already strong evidence, that the bulk model is not able to explain the phenomenology of our observed overshoot effect. We find the overshoot effect to be stronger if the electrochemical potential $\mu^*$ resides in a comparatively large Landau gap rather than a spin gap in the bulk, for the lowest temperatures and the highest electron densities. In all three cases, inter-level scattering or level-coupling is suppressed in the bulk such that the basic ingredients of the bulk model are not fulfilled. 
An alternative explanation for the overshoot is provided by the edge model which is based on the Landauer-B\"uttiker picture of the IQHE. In order to explain the overshoot effect, the edge model makes a number of assumptions that cannot be applied to our system. It e.g. relies on a strong spin-orbit interaction as a mechanism which couples both spin states at the sample edges \cite{KOM1991}.

In our experiments, we observe the overshoot anomaly to occur at the low magnetic field end of a Hall plateau. In this regime, the current distribution is still confined to an IS in the framework of the screening theory. However, the current distribution spreads out to the next local resistance minimum for decreasing magnetic fields as the IS becomes evanescent and breaks down. Hence, the occurrence of the overshoot effect appears to be related to the breakdown of the current carrying IS.

\begin{figure}
\begin{center}
\includegraphics[width=1\columnwidth]{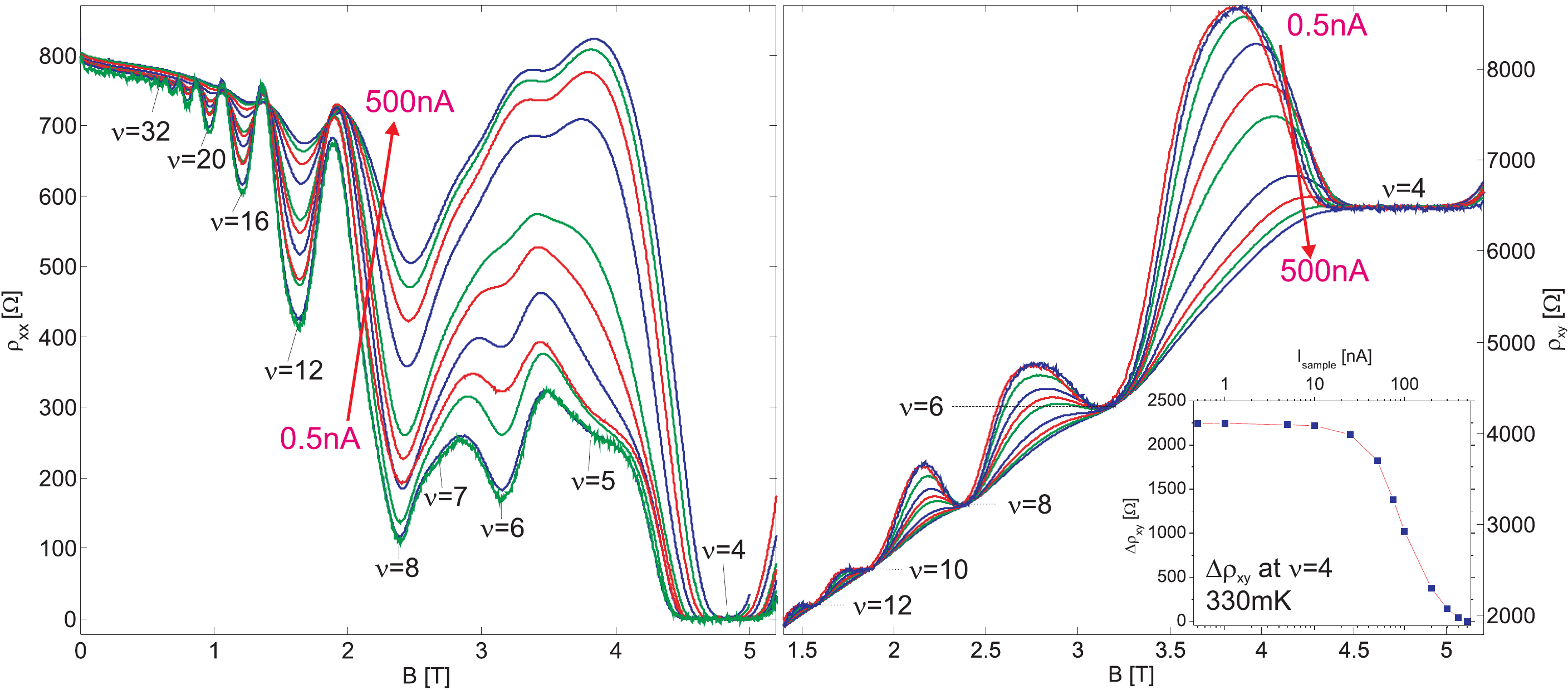}
\end{center}

\caption{(Color online) Excitation current dependence of the overshoot at $T=\mathrm{330~mK}$. The overshoot can be selectively destroyed by increasing the excitation current. Inset: Magnitude of the overshoot as a function of excitation current. The overshoot decreases with increasing excitation current.}

\label{I-series}
\end{figure}

To examine this indication in further detail, we studied the effect of imposed excitation current over a wide range of amplitudes. According to calculations \cite{SiddikiEPL:09}, the sample current was found to predominantly affect the width and stability of incompressible edge states which has consequences on the breakdown of the current carrying IS. Figure~\ref{I-series} shows a current series measured with a $\mathrm{20~{\mu}m}$ wide Hall bar at a bath temperature of $\mathrm{330~mK}$. The sample is illuminated upon saturation of the sheet carrier density and the density is kept constant for the entire measurement series at $n_\mathrm{2DEG}=\mathrm{4.45\times10^{11}~cm^{-2}}$. The current is varied by three orders of magnitude from $\mathrm{0.5~nA}$ to $\mathrm{500~nA}$. The left panel of figure~\ref{I-series} shows $\rho_\mathrm{xx}$, while the right side presents a zoom-in of the simultaneously recorded $\rho_\mathrm{xy}$.
For the lowest sample currents, Shubnikov-de~Haas oscillations with a 4-fold periodicity start around $B\approx \mathrm{0.6~T}$ for ${\nu}=\mathrm{32}$. Above $B\approx\mathrm{2~T}$ Zeeman- (${\nu}=\mathrm{6}$) and also valley-splittings (${\nu}=\mathrm{5},~\mathrm{7}$) become resolvable. When increasing the current, the amplitudes of the oscillation in $\rho_\mathrm{xx}$ decrease and are less pronounced, especially for higher filling factors.
In $\rho_\mathrm{xy}$, filling factor ${\nu}=\mathrm{12}$ is resolved first and strong overshoots are observed at filling factors ${\nu}=\mathrm{8},~\mathrm{6}$~and~$\mathrm{4}$. These overshoots decrease rapidly with increasing current. The inset in the lower right side of figure~\ref{I-series} plots $\Delta\rho_\mathrm{xy}$ for filling factor ${\nu}=\mathrm{4}$ exemplarily versus applied excitation current. Starting at low currents, the magnitude of the overshoot is almost unaffected up to approximately $\mathrm{10~nA}$. Increasing the current further leads to a rapid decrease of the magnitude of the overshoot until complete suppression at $\mathrm{500~nA}$. However, at $\mathrm{500~nA}$ the quantized Hall plateau is still existent in $\rho_\mathrm{xy}$. That is, by increasing the current, the overshoot can be completely suppressed while the quantum Hall effect is preserved and proves to be much more stable than the overshoot.

\begin{figure}
\begin{center}
\includegraphics[width=0.6\columnwidth]{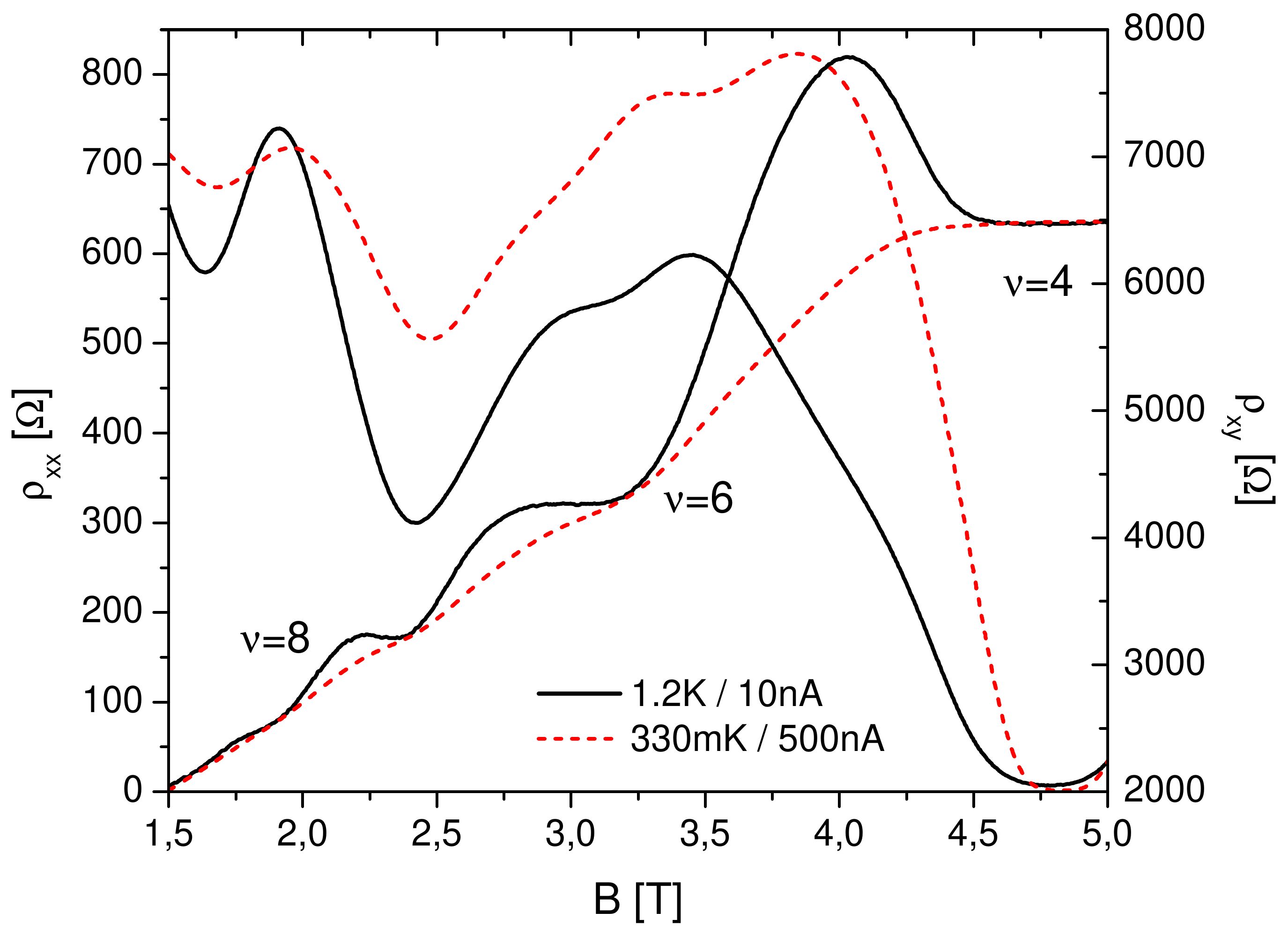}
\end{center}

\caption{(Color online) Direct comparison of the effect of increasing sample current and increasing temperature. At a lattice temperature of $T=\mathrm{330~mK}$ and $\mathrm{500~nA}$ sample current the overshoot is completely suppressed, still preserving $\rho_\mathrm{xx}=\mathrm{0}$, while at $T=\mathrm{1.2~K}$ but only $\mathrm{10~nA}$ sample current the overshoot is still very prominent but the $\rho_\mathrm{xx}=\mathrm{0}$ minimum is already lifted.}

\label{compItoT}
\end{figure}

To rule out heating of the entire 2DEG as the reason for the suppression of the overshoot, we compare a measurement of $\rho_\mathrm{xy}$ and $\rho_\mathrm{xx}$ at a low bath temperature of $\mathrm{330~mK}$ and a high sample current of $\mathrm{500~nA}$ (red dotted line) exemplarily against a measurement at a high temperature of $\mathrm{1.2~K}$ but a low current of $\mathrm{10~nA}$ (black solid line) in figure~\ref{compItoT}. 
At $T=\mathrm{1.2~K}$, $\rho_\mathrm{xy}$ shows signatures of the quantum Hall effect with a prominent overshoot at filling factor 4. The Shubnikov-de Haas oscillations in $\rho_\mathrm{xx}$ are pronounced, however, do not reach $\rho_\mathrm{xx}=\mathrm{0~\Omega}$ for any filling factor. In comparison, $\rho_\mathrm{xx}$ at $T=\mathrm{330~mK}$ but $\mathrm{500~nA}$ does reach $\rho_\mathrm{xx}=\mathrm{0~\Omega}$ at filling factor 4. In $\rho_\mathrm{xy}$, however, the overshoot is completely suppressed. Consequently, a high lattice temperature of $T=\mathrm{1.2~K}$ lifts the $\rho_\mathrm{xx}=\mathrm{0~\Omega}$ minimum but preserves the overshoot, whereas a high current of $I=\mathrm{500~nA}$ completely suppresses the overshoot and preserves the extended $\rho_\mathrm{xx}=\mathrm{0~\Omega}$ minimum. Hence we conclude, that the selective suppression of the overshoot cannot be explained by a current induced 2DEG heating effect alone.

By increasing the current $I$ with fixed $w_{\mathrm{bar}}$, the bulk current density $j=I/w_{\mathrm{bar}}$ is increased, as well. In order to be able to differentiate whether the current density $j$ or the current $I$ itself are essential for the suppression of the overshoot or not, a further experiment is necessary. If the overshoot phenomenon was caused by bulk contributions, the overshoot is expected to behave similarly, no matter how the current density $j=I/w_{\mathrm{bar}}$ is increased. To discriminate between the effects of increasing the current density $j$ by increasing $I$ or reducing $w_{\mathrm{bar}}$, we designed a Hall bar with different sections where $w_{\mathrm{bar}}$ is varied in steps from $\mathrm{200~{\mu}m}$ to $\mathrm{25~{\mu}m}$.
We were thus able to take data for each $w_{\mathrm{bar}}$ and current $I$ simultaneously on one sample under exactly equal conditions in a single cool-down.

\begin{figure}
\begin{center}
\includegraphics[width=0.55\columnwidth]{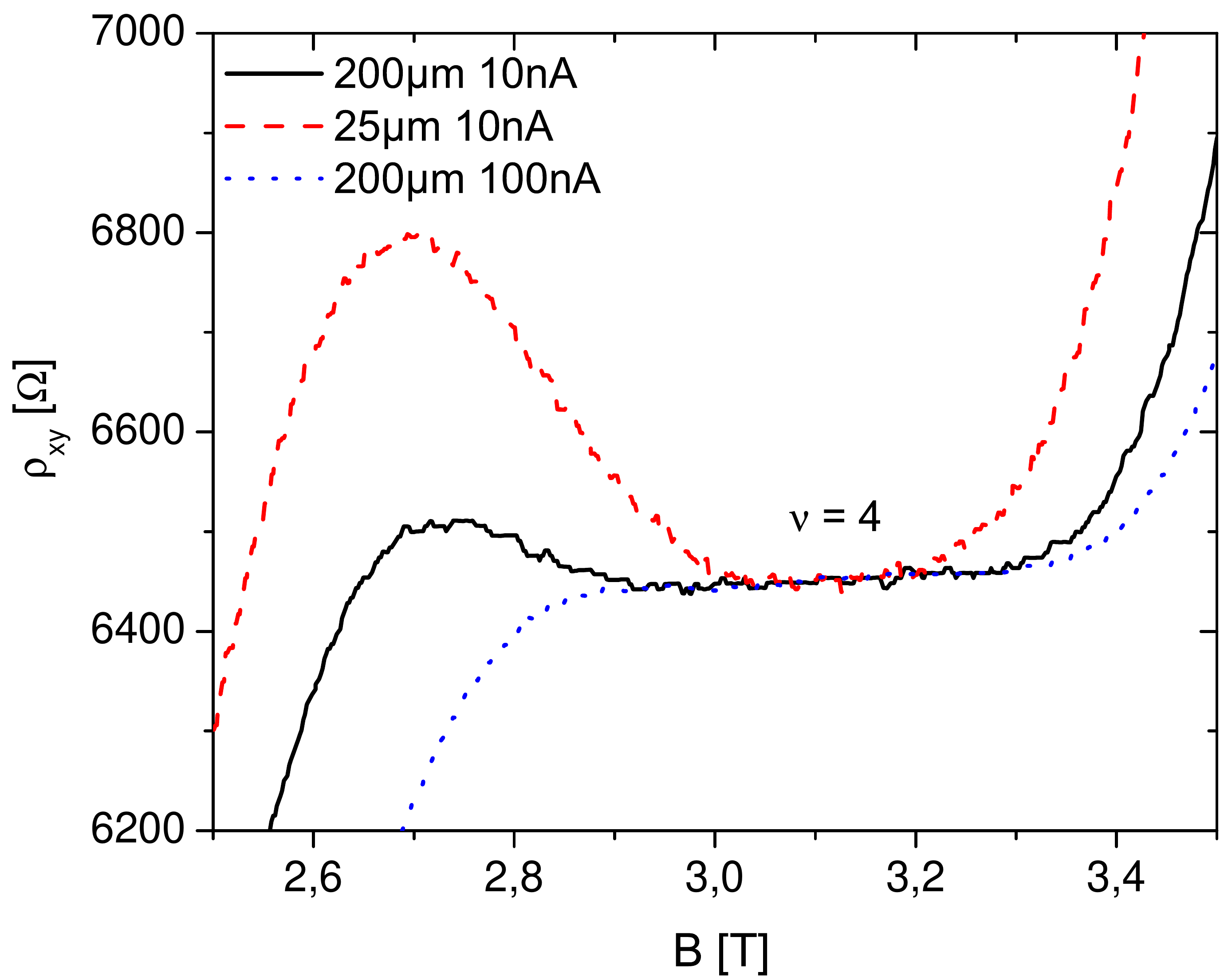}
\end{center}

\caption{(Color online) Bulk current density dependence of the overshoot. Increasing the current density by increasing the excitation current or by decreasing $w_{\mathrm{bar}}$ yields opposite results. The first suppresses the overshoot, the latter increases the overshoot.}

\label{currentdensity-series}
\end{figure}

Starting at a low current density $j$, figure~\ref{currentdensity-series} shows an overshoot for the $\mathrm{200~{\mu}m}$ wide Hall bar section at an excitation current of $\mathrm{10~nA}$. Increasing the current density $j$ by a factor of 8 by decreasing $w_{\mathrm{bar}}$ to $\mathrm{25~{\mu}m}$ and keeping $I=\mathrm{10~nA}$ fixed leads to an increased overshoot. In contrast, increasing the current density $j$ by a factor of 10 by increasing the current to $I=\mathrm{100~nA}$ and keeping $w_{\mathrm{bar}}=\mathrm{200~{\mu}m}$ fixed suppresses the overshoot. As a similar increase of $j$ leads to opposite results regarding the overshoot, the averaged bulk current density $j=I/w_{\mathrm{bar}}$ is no relevant figure of merit to describe the overshoot phenomenon. Instead, only the current $I$ itself, which is crucial for the current density in the incompressible edge state, determines whether the overshoot is suppressed or not. Analogously, these measurements of the increasing overshoot effect with decreasing sample size also demonstrates that a lower contribution of the bulk of the 2DEG to electron transport is essential for a pronounced overshoot. Hence, we conclude that the overshoot phenomenon originates from the edge state transport regime.

\section{Discussion}
\label{sec:L04_Discussion}

In section~\ref{sec:L02_ScreeningBasics} and section~\ref{sec:L03_Experiments}, we only discussed the IQHE and the overshoot phenomenon considering one existing or vanishing IS. Current was either flowing in one IS or increasingly in the bulk as an IS breaks down.

The imposed sample current $I_0$ into a Hall bar extending from $-d$ to $d$ across the bar can be related to the current density along the Hall bar $j_y(x)$ and by the resistivity tensor to the electrochemical potential in the quantized regime via
\begin{equation}
I_0=\int^{+d}_{-d} dx~j_y(x)=\frac{e}{h}\int^{+d}_{-d}dx~\nu(x)\frac{\partial \mu^*}{\partial x}.
\label{eq:eqI0}
\end{equation}
In equation~\ref{eq:eqI0}, the local resistivity tensor components have been introduced in $j_y(x)=\mathrm{1}/\rho_\mathrm{xx}(x)E_y(x)=\mathrm{1}/\rho_\mathrm{xy}(x)E_x(x)$, together with $\rho_\mathrm{xy}(x)=h/\nu(x)e^2$ and $E_x(x)=e^{-1}{\partial}\mu^*(x,y)/{\partial}x$. This description will be best towards zero temperature when the drop of  $\mu^*(x,y)$ across an IS of width $W_N$ can be approximated by a constant transverse electric field $E_x$ comparable to the schematical picture of figure~\ref{fig:01_ScreeningBasics}(b). After integration over both branches of an IS with local filling factor $\nu(x)=N$ at the two edges of the Hall bar, one obtains for the Hall bar current
\begin{equation}
I_0=\mathrm{2}\frac{e}{h}e{E_x^N}N{W_N}.
\label{eq:eqI0N}
\end{equation}
This allows to calculate the measured Hall voltage $V_{\mathrm{Hall}}=\mathrm{2}{W_N}E_x^N$. We find the familiar globally measured Hall resistance for the Hall plateau of filling factor~$N$
\begin{equation}
\rho_\mathrm{xy}^N=\frac{V_\mathrm{Hall}}{I_0}=\frac{h}{{e^2}N}.
\label{eq:rhoxyN}
\end{equation}
From this simple consideration we can already conclude that as one IS becomes evanescent, the current density $j_y(x)$ and thus ${\partial\mu^{*}(x,y)}/{\partial x}$ will be reduced within an IS region and consequently the Hall resistance must drop. Conversely, more than one IS is required to explain anomalies such as the overshoot, where the drop of the chemical potential across the Hall bar needs to rise in total.

In the following, we will first show that the co-existence of two evanescent IS leads to the overshoot effect and that this model explains our observations from the previous section.

\subsection{Co-existence of Incompressible Strips}

\begin{figure}
\begin{center}
\includegraphics[width=1\columnwidth]{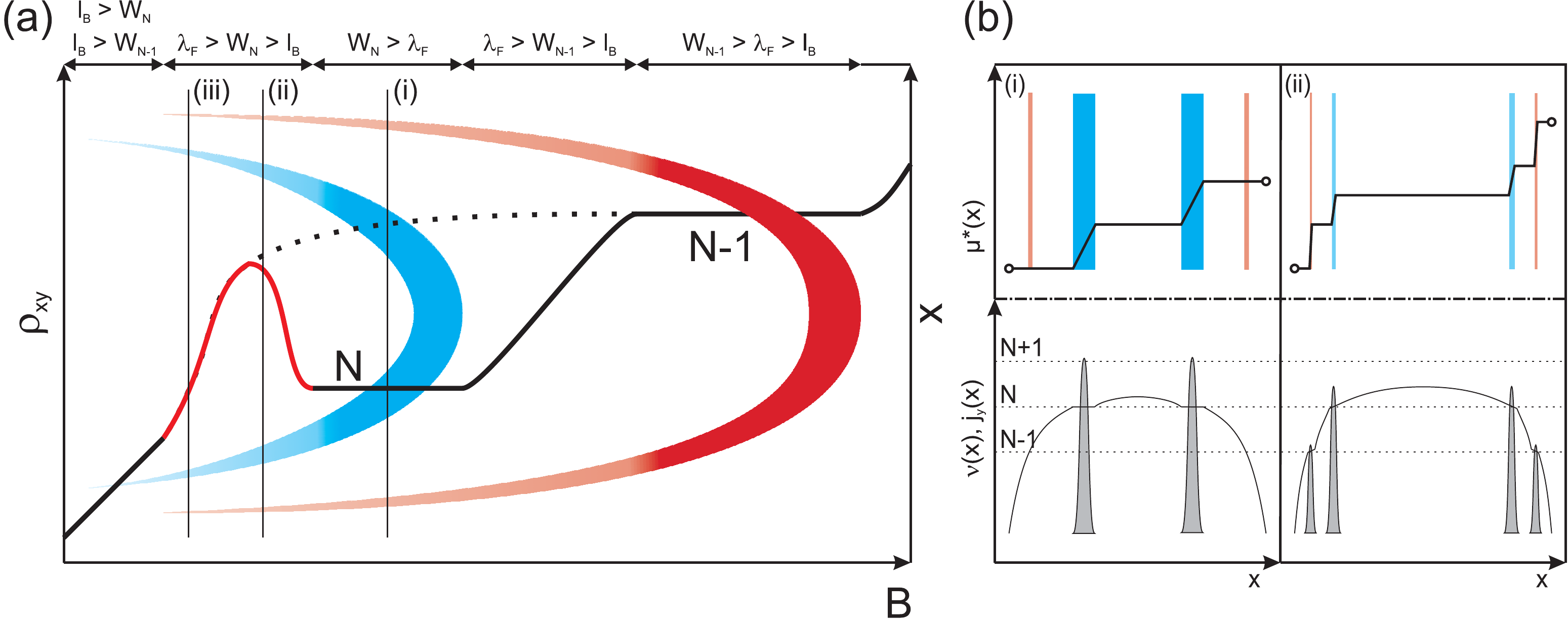}
\end{center}

\caption{(Color online) (a) Formation of overshoot as a result of co-existing evanescent IS at the low magnetic field end of the filling factor $N$ Hall plateau. (b) Upper panel: Course of the chemical potential for cross-sections (i) and (ii) in (a). Lower panel: Current distribution for cross-sections (i) and (ii). For cross-section (i), despite the presence of an evanescent IS of filling factor $N-1$, current is fully confined to the incompressible region of filling factor $N$ and the Hall resistance is quantized. In (ii), the co-existing evanescent IS of filling factors $N$ and $N-1$ constitute local resistance minima in comparison to the compressible bulk which is why the current is confined to them. This leads to a stronger transverse electric field $v_y\propto\partial_x\mu^*(x,y)\propto{E_x}$ and thus a larger Hall voltage is generated. }

\label{fig:05_Coexistence}
\end{figure}

Figure~\ref{fig:05_Coexistence}(a) sketches the Hall resistance together with IS regions including the overshoot phenomenon at filling factor $N$.
At the low magnetic field end of the $\rho_\mathrm{xy}$ plateau at filling factor $N-1$, the light colored (red) IS becomes evanescent when $l_B<W_{N-1}<\lambda_\mathrm{F}$ and $\rho_\mathrm{xy}$ drops. In this situation, $l_B$ is sufficiently small so that the next bulk IS of filling factor $N$ forms before the evanescent strip of filling factor $N-1$ close to the sample edges breaks down completely. This configuration is shown in cross-section in (i) of figure~\ref{fig:05_Coexistence}(b). In the upper panel, we show the course of $\mu^*$ across the Hall bar, whereas the lower panel depicts the corresponding current distribution $j_y(x)$ and the local filling factor $\nu(x)$. In the regime of $W_{N}>\lambda_\mathrm{F}$, $\mu^*$ only drops in the well developed inner IS with local filling factor $N$ to which the imposed current is confined. 

Cross-section (ii) illustrates the situation in an overshoot regime. Now $l_B<(W_{N-1},W_{N})<\lambda_\mathrm{F}$ and the IS of filling factor $N$ becomes evanescent. This leads to current leaking out of this IS. However, in addition, the evanescent IS of filling factor $N-1$ is also still present which constitutes a local resistance minimum compared to the surrounding compressible region. Hence, current escapes from the IS of filling factor $N$ to the IS of filling factor $N-1$.
From equation~\ref{eq:eqI0}, we can derive an expression similar to equation~\ref{eq:eqI0N} for the situation when the imposed current redistributes over two evanescent incompressible strips with filling factors $N$ and $N-1$.
\begin{equation}
I_0=I_N+I_{N-1}=\mathrm{2}\frac{e^2}{h}(E_x^{'N}N{W_N^{'}}+E_x^{N-1}(N-1)W_{N-1}).
\label{eq:eqI0NN-1}
\end{equation}
The Hall voltage will essentially only be created by the drop of $\mu^*(x,y)$ across the different IS, hence
\begin{eqnarray*}
V_\mathrm{Hall} &=&\mathrm{2}W_N^{'}E_x^{'N}+\mathrm{2}W_{N-1}E_x^{N-1}\\
    &=&h/e^2(I_N/N+I_{N-1}/(N-1)).
\end{eqnarray*}
By applying current conservation $I_N=I_0-I_{N-1}$, one can derive an approximation for the Hall resistance in the overshoot regime.
\begin{equation}
\rho_\mathrm{xy}^N=\frac{h}{e^2}\left(\frac{1}{N}+\frac{I_{N-1}}{I_0}\left(\frac{1}{N-1}-\frac{1}{N}\right)\right).
\label{eq:rhoxyNN-1}
\end{equation}
Finally, in the regime of cross-section (iii), one or both IS $N-1$ and $N$ become smaller than the extent of the wavefunction and the Hall resistance drops toward the next Hall plateau.

Equation~\ref{eq:rhoxyNN-1} contains the overshooting character of the Hall resistance: Since current redistributes from an IS with a larger filling factor $N$ to one with a smaller number of levels $N-1$, electrons have to be accelerated to conform to current conservation. Hence the entire drop $\mu^*(x,y)$ across the two IS increases according to $v_y(x)\propto{\partial}\mu^*(x,y)/{\partial}x$. Of course, the exact relative current $I_{N-1}/I_0<1$ will depend on local characteristics of the IS, but the deviation of the Hall resistance $\rho_\mathrm{xy}^N$ to the Hall plateau value is always positive.

This simple model demonstrates that an overshoot effect of the Hall voltage will occur whenever co-existing evanescent IS are available. This is only possible if, for a given magnetic field range, $\mathrm{l_B<W_{N-1},W_{N}<\lambda_\mathrm{F}}$ holds. This peculiar configuration cannot be expected to be realized under all experimental conditions and thus provides a means of testing the above model by comparing the predictions of the overshoot model to a measured sequence of overshoots and to the behavior of the overshoot effect with varying experimental parameters.

\subsection{Comparison with Experiment}

The basic treatment of co-existing evanescent IS will now be extended in order to explain all details of the overshoot effect observed in the section~\ref{sec:L03_Experiments}.

\begin{figure}
\begin{center}
\includegraphics[width=1\columnwidth]{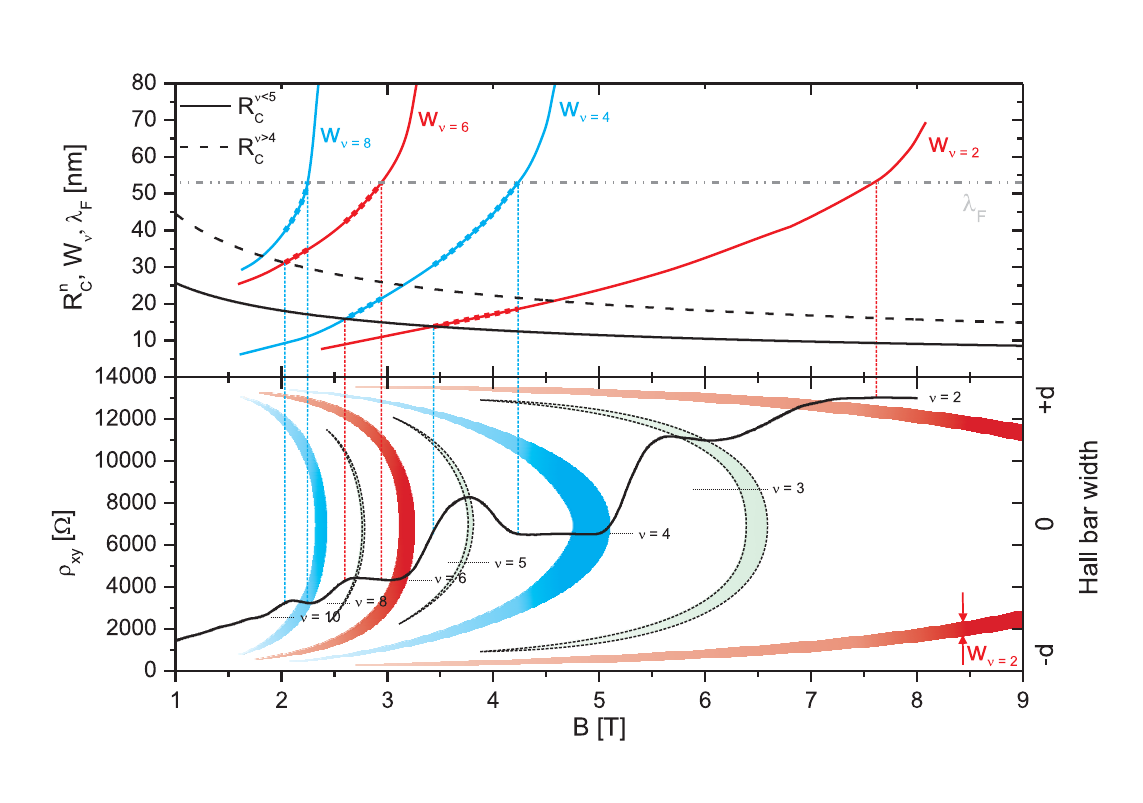}
\end{center}

\caption{(Color online) The lower panel depicts $\rho_\mathrm{xy}$ for a typical sample with $n_\mathrm{2DEG}=\mathrm{4.46\times10^{11}~cm^{-2}}$ at $T=\mathrm{320~mK}$. The onset and end of Hall plateaus and overshoots are used to reconstruct the schematic course of incompressible regions for different filling factors. These crescent-like IS regions are superimposed to the data.
The width $W_{\nu}$ of different IS (red and blue solid lines), the Fermi wavelength $\lambda_\mathrm{F}$ (horizontal gray dash-dotted line) and the extent of the wave function for different Landau levels $R_C^n$ (black solid and black dashed lines) are plotted in the upper panel. As long as $\lambda_\mathrm{F}>W_{\nu}>R_C^n$, the IS associated with filling factor $\nu$ is evanescent and current is partially confined to this region before the IS breaks down. An overshoot forms only in a regime of two co-existing evanescent IS, marked with vertical dotted lines of equal color, as \textit{e.g.} around $B=\mathrm{3.8~T}$. For higher magnetic fields current is completely confined to the well developed IS of filling factor $\nu=\mathrm{4}$ whereas for lower magnetic fields the already evanescent IS of filling factor $\nu=\mathrm{2}$ breaks down completely as it finally becomes smaller than $R_C^n$.}

\label{fig:05_Coex_in_SiGe}
\end{figure}

The lower panel of figure~\ref{fig:05_Coex_in_SiGe} shows an exemplary Hall resistance for a sample with $n_\mathrm{2DEG}=\mathrm{4.46\times10^{11}~cm^{-2}}$ at a bath temperature of $T=\mathrm{320~mK}$. Starting at high magnetic fields, the evolution of different IS regions associated with different filling factors can be reconstructed from the onset and end of both quantum Hall plateaus and overshoot regions, as shown. Landau (blue), spin (red) or valley gap (green) associated IS regions are given different colors. IS regions related to valley gaps are only adumbrated since such gaps cannot be resolved experimentally in the considered magnetic field range. 

The upper panel of figure~\ref{fig:05_Coex_in_SiGe} plots the cyclotron radius $R_C^{n}=l_B\sqrt{2n-1}$ \cite{Beenakker2004} for different Landau levels in black (solid and dashed black lines). It is a measure for the extent of the respective wave function, with the Landau level index $n$ and the magnetic length $l_B$. The gray, dash-dotted horizontal line represents the Fermi wavelength $\lambda_\mathrm{F}=\sqrt{\mathrm{4}\pi/n_\mathrm{2DEG}}$ of the sample. The approximate evolution of IS widths $W_{\nu}$ is given in the same colors in the upper part of figure~\ref{fig:05_Coex_in_SiGe} as the corresponding IS regions in the lower panel of figure~\ref{fig:05_Coex_in_SiGe}. During the existence of a Hall plateau of filling factor $\nu$, $W_\nu>\lambda_\mathrm{F}$. At the low magnetic field end of a Hall plateau, the width of the respective IS is $W_\nu\approx\lambda_\mathrm{F}$. Both, $R_B^n$ and $\lambda_\mathrm{F}$, define a window in which IS can co-exist, become evanescent and carry current before a respective IS breaks down. Vertical dotted lines mark magnetic field intervals in which at least two evanescent IS co-exist as $R_C^n<W_\nu<\lambda_\mathrm{F}$ holds for each of them.

As soon as an IS forms in the bulk for high magnetic fields, a quantum Hall plateau appears. When $W_{\nu}$ of a certain IS enters the evanescence window, $W_{\nu}<\lambda_\mathrm{F}$, the corresponding Hall plateau breaks down either by $\rho_\mathrm{xy}$ dropping to the next plateau, as for filling factor $\nu=\mathrm{2}$ in the lower panel of figure~\ref{fig:05_Coex_in_SiGe}, or by forming an overshoot. Several of such situations are realized whenever $W_\nu<\lambda_\mathrm{F}$ until $W_{\nu-2}<R_C^n$. These co-existence regimes are highlighted by solid-dotted sections of the IS widths in the upper panel of figure~\ref{fig:05_Coex_in_SiGe}. Thus, the overshoot at the low magnetic field end of the $\rho_\mathrm{xy}$ plateau of filling factor, \textit{e.g.} $\nu=\mathrm{4}$, is caused by co-existing evanescent edge IS of filling factors $\nu=\mathrm{2}$ and $\nu=\mathrm{4}$. 

We only considered the co-existence of evanescent IS edge states stemming from even filling factors so far. As shown in the lower panel of figure~\ref{fig:05_Coex_in_SiGe}, a not fully developed bulk IS can exist at the position of the overshoot, e.g. around $B=\mathrm{3.8~T}$. In fact, such a nearly bulk incompressible region could act as a local resistance minimum that leads to a partial redistribution of the current to this spatial region \cite{Mares:09}. In such a case, the co-existence of not only two but three evanescent IS regions of three different filling factors give rise to the overshoot effect.

In a next step, we can now go beyond the static view of figure~\ref{fig:05_Coex_in_SiGe} which explains the presence or absence of overshoots for only one set of experimental parameters.
The above considerations can be combined with the experimental findings on the stability of overshoots under different conditions.
In the overshoot regime of a given filling factor, both contributing evanescent IS can be expected to be of different width. The outer IS tends to be more fragile and is thus closer to breakdown than the inner IS. Consequently, by varying parameters that affect the stability of a single IS, both current carrying IS  should be influenced. But the more fragile IS is more easily destroyed. In this situation, the usual IQHE is recovered. Such a suppression of the overshoot phenomenon is demonstrated in the temperature, density and current(-density) series in section~\ref{sec:L03_Experiments}.

The temperature dependence of the overshoot in figure~\ref{T-series} corresponds to a temperature induced breakdown of the outer and more fragile IS \cite{Lier94:7757,Oh97:13519}. Only one evanescent IS remains, explaining the reduced overshoot with increasing temperature.

The observed increase of the overshoot with 2DEG density in figure~\ref{N-series} can be explained by two different effects: First, screening of the disorder is improved and thus the probability of electron scattering and tunneling through an IS is reduced \cite{Gulebaglan}. Second, a higher 2DEG density shifts the position of each filling factor or IS to higher magnetic fields and thus higher energy gaps. Both effects contribute to more stable IS in the evanescence window and thus to a stronger overshoot effect.

The effect of increasing sample current in figure~\ref{I-series} is also two-fold: in the evanescence window, the electron gas will be heated locally when electrons start to leak out of an IS and scatter in the surrounding compressible region at the low magnetic field end of a Hall plateau. The more fragile, outer IS will again be affected first. The overshoot is destroyed with increasing sample current. Nevertheless the inner IS is less affected leading to a preserved Hall plateau. The second effect of high sample currents is a tilting of the potential landscape \cite{SiddikiEPL:09} in the out-of-linear-response regime. Both evanescent IS at one edge boundary become wider on the expense of both IS at the edge boundary on the other side. For a certain current amplitude the narrowest outer IS breaks down, resulting in a breakdown of the overshoot.

In all three experiments a stronger outer IS should of course correlate with a more robust inner IS. In agreement with theoretical predictions, we observe that an increasing overshoot is accompanied by a wider extension of the quantized Hall plateau on the magnetic field axis as well as a shift of the overshoot maximum towards higher filling factors. Both observations are evidence for the increased stability of the inner IS.

Finally, we can explain the size dependence of the overshoot, which increases with smaller Hall bar width as evidenced in figure~\ref{currentdensity-series}. At the low magnetic field end of a Hall plateau, current starts to leak out of the inner IS and redistributes to the next resistance minimum. For a smaller sample size, less current flows in the bulk, but in the resistance minimum of the other evanescent IS leading to an increased magnitude of the overshoot.

By comparing the relative overshoot strengths in figure~\ref{Intro}, Landau gap associated overshoots (${\nu=\mathrm{4}}, \mathrm{8}, \mathrm{12}, \dots$) were found to be comparatively strong with respect to spin gap associated overshoots (${\nu=\mathrm{6}}, \mathrm{10}, \dots$). This pattern indicates the specific nature of the energy gaps to be strongly related to the exact stability of the outer evanescent IS. The spin polarization of the 2DEG follows an analogous pattern: spin polarization for filling factors ${\nu=\mathrm{6}}, \mathrm{10}$ is extremal, whereas it is zero for ${\nu=\mathrm{4}}, \mathrm{8}, \mathrm{12}$. This spin polarization can lead to exchange enhanced energy gaps depending on the local filling factor \cite{Manolescu1995} within the 2DEG. Such effects studied in detail, could explain nuances in the relative overshoot magnitude for different underlying energy gaps, but have not been treated numerically in a self-consistent fashion for the given system yet.

\subsection{Overshoot in other 2DEG Host Systems}

The employed model relying on co-existing evanescent IS, is universal to explain all aspects of the overshoot phenomenon in our Hall bar devices but does not depend on peculiarities of the examined material system. A comparison of the presented results with overshoot studies in other host systems shows that overshoots are especially strong and occur for many filling factors in Si/SiGe and Si-MOS structures and are less pronounced in other material systems such as GaAs. When extending the scope of the overshoot discussion to such systems, three properties become important: the Fermi wavelength, the energy level structure which is influenced by the effective mass and the effective \emph{g}-factor and the confinement potential at the edges of the Hall bar.

The Fermi wavelength in Si is a factor of $\sqrt{2}$ larger than in GaAs for the same electron density, due to the valley degeneracy factor. For this reason, the evanescence window in GaAs, in which at least two IS can co-exist, is smaller. Thus overshoots are more unlikely to occur or are even suppressed.

The effective mass and g-factor determine the relative gap sizes of Landau and spin gaps. In GaAs, for example, $m^*$ is about a factor of three smaller than in Si, and similarly the effective g-factor. This leads to a larger Landau level splitting and a reduced Zeeman splitting in GaAs compared to Si. Combined with the typically much lower scattering rates in GaAs, IS associated with Landau gaps are more robust compared to spin-associated IS, or all types of IS in Si. As a result, it will be more likely to find an evanescent co-existing IS of a Landau gap at the low magnetic field end of a spin gap-associated Hall plateau (odd filling factors) than a stable IS of a spin gap at the low field end of a Landau gap Hall plateau (even filling factors). Indeed, overshoots in GaAs have only been found at odd, i.e. spin gap related, filling factors \cite{KOM1991,RIC1992,KOM1993}.

The shape of the confinement potential of the 2DEG is predominantly influenced by the technique used to realize a Hall bar device. Common etch-defined Hall bar devices, as used in this work, usually present a steep confining potential whereas gate-defined devices present a smoother but also adjustable confinement. A steeper confinement shrinks the width of IS \cite{Hor2008} thereby reducing the evanescent window such that overshoots will be smaller or not even present. Thus, the realization and investigation of gate-defined Hall bar devices, in Si and especially other host systems, opens up a perspective to directly test the influence of the confinement potential and the predictions of our model.

\subsection{Conclusion}

In this work we have described an anomalous behavior of the IQHE that is referred to as quantum Hall resistance overshoot, both experimentally and theoretically. The quantum Hall resistance overshoot has been frequently observed in Hall devices realized in different material systems, but the various characteristics of the overshoot have not been examined in detail so far and could not be explained consistently and independently of the specific properties of a respective material system.

Here, we investigated the nature of the quantum Hall resistance overshoot effect based on its behavior under different experimental parameters such as the temperature, the 2D sheet carrier density, the sample current and the sample geometry.
To model our experimental findings, we used a theoretical approach based on the self-consistent screening theory for the IQHE that considers direct Coulomb interaction within the 2DEG. 
The possibility to describe all aspects of the overshoot phenomenon demonstrates that direct Coulomb interaction within a 2DEG is crucial for understanding many aspects of the IQHE and the current distribution in the quantized regime within a sample. 
Within the screening theory, current confinement to one incompressible region gives rise to the IQHE. Our work illustrates the effect of current confinement to more than one incompressible region for the first time. The overshoot phenomenon turns out to be a natural outcome of current confinement to co-existing evanescent incompressible strips of different filling factors at the sample edges. 
This model is capable to explain all details of our experiments and can also be extended to other material systems easily.

\begin{acknowledgments}

We acknowledge financial support of the Deutsche Forschungsgemeinschaft via the Sonderforschungsbereich 631, Teilprojekt C4 and stimulating discussions with G.~Abstreiter and M.~Grayson. A.S. is partially supported by TUBiTAK under grant no:109T083.

\end{acknowledgments}

\bibliographystyle{apsrev4-1}
%

\end{document}